\documentclass[preprint,showpacs,preprintnumbers,amsmath,amssymb]{revtex4}
\usepackage{graphicx}% Include figure files
\usepackage{graphics}% Include figure files
\usepackage{dcolumn}% Align table columns on decimal point
\usepackage{bm}% bold math
\begin{document}

\title{Polarization entanglement visibility of photon pairs emitted 
by a quantum dot embedded in a microcavity}

\author{J.I. Perea}
\affiliation{Departamento de F\'{\i}sica Te\'orica de la Materia Condensada,
Universidad Aut\'onoma de Madrid, 28049 Madrid, Spain.}

\author{C. Tejedor}
\affiliation{Departamento de F\'{\i}sica Te\'orica de la Materia Condensada,
Universidad Aut\'onoma de Madrid, 28049 Madrid, Spain.}

\begin{abstract}
We study the photon emission from a quantum dot embedded in a microcavity. 
Incoherent pumping of its excitons and biexciton provokes the emission  
of leaky and cavity modes. By solving a master equation we obtain the 
correlation functions required to compute the spectrum and the 
relative efficiency among the emission of pairs and single photons.
A quantum regime appears for low pumping and large rate of emission.  
By means of a post-selection process, a two beams experiment with different 
linear polarizations could be performed producing a large polarization 
entanglement visibility precisely in the quantum regime.
\end{abstract}

\maketitle

\section{Introduction}
A quantum dot (QD) embedded in a microcavity (either a pillar or a 
photonic crystal) can be an efficient emitter 
of photons when the excitation of an electron-hole pair 
(usually labeled as an exciton) is close to resonance with an optical 
mode of the cavity \cite{gerard,michler,moreau,santori,pelton,benyoucef}. 
The strong coupling regime between excitons and cavity photons has been 
recently shown \cite{reithmaier,yoshie,peter} which opens a set of 
possibilities for using this system in quantum information protocols. 
Up to now, many efforts have been devoted to the controlled emission of 
single photons. A subsequent manipulation of such single photons has allowed 
to establish they were indistinguishable by detecting the correlation between 
photon pairs \cite{santori}. This work has a different target. 
We are interested in studying the possibility of having a source of 
efficient emission of photon pairs in which quantum information could
be stored. For this purpose we consider a neutral QD in a cavity with 
precise conditions. The lowest part of the QD spectrum is formed by two 
degenerate (or almost) excitons and one biexciton. 
One expects the photon pair emission to be favored with respect to 
that of single photons when there is not any cavity photon in resonance 
with any single exciton but the energy of the biexciton with respect to 
the ground state is exactly (or very close to) twice the cavity photon energy.
This spectrum makes the QD inside the cavity an excellent candidate as 
an efficient source of pairs of photons. Moreover, it has the added value 
of the possible manipulation of the two polarizations of the cavity 
photons as will be discussed carefully below.    

We have performed a theoretical analysis of the photon emission from the 
system described above in the regime in which cavity modes and QD 
excitations are strongly coupled to each other. 
The system is also coupled to the outside which has a number of degrees 
of freedom so large that can be considered
as a reservoir. Three are the processes of interest: the scape of cavity 
modes from the cavity, the emission of leaky modes due to transitions in
the QD and the incoherent pumping of the QD. The magnitudes of interest 
for describing photon emission are obtained from
two-times correlation functions or their Fourier transforms \cite{walls}.
Tracing out in the external degrees of freedom, one can get the time 
evolution of the density matrix $\rho $ of just the QD and the cavity. 
By using the quantum regression theorem, correlation functions are obtained 
from the dynamics of the density matrix. 
Our study produces two main results: 

First, it turns out that the simple picture of looking for the two-photon 
resonance described above is not describing properly the physics of 
these systems. The understanding of the actual mechanisms allows us to 
determine the quantum regime of interest for photon pair emission. 

Our second important result concerns to the manipulation of the 
correlation between the two emitted photons. Since cavity photons can 
be degenerated (or almost) depending on the shape of the electric and 
magnetic components \cite{panzarini,yamamoto02}, one has the possibility 
of getting correlation between the polarizations of the two emitted photons. 
This can be addressed by making a Hambury-Brown Twiss-like\cite{walls} 
or a two-photon interference \cite{hong,santori} experiment with 
linearly polarized photons. An important aspect to be pointed out is that  
the possible entanglement of the photons forming a pair requires two degrees 
of freedom: The first is, obviously, the photon polarization.
The second degree of freedom could be associated with two different
frequencies of the photons, but a more interesting alternative is the 
splitting of the photons in two different beams. 
To our knowledge, no QD inside cavities emitting in two preferential
directions have been fabricated. Therefore, we focus on a less ambitious target
and invoke a post-selection procedure by means of a non-polarizing beam
splitter. In a probabilistic way, only half of the emitted pairs would have 
the required double degree of freedom of polarization and direction. 
After the separation in two beams, these pairs could be used for
experiments similar to the ones,  
performed in other systems\cite{ou88,kwiat99,volz01,zwiller02}, measuring 
different linear polarizations on each of the beams.
Having in mind this fact, our figure of merit is the entanglement visibility, 
precisely defined below, which gives insight about the availability of 
storing information in the relative angle between the linear polarizations 
of the two photons.

The paper is organized as follows: section II presents our theoretical 
framework while section III contains the results and a discussion of them.

\section{Theoretical framework}
We consider just four levels of the QD:
the ground state $G$, the two excitons $X^+$, $X^-$ with third component 
of their angular momentum equal to $\pm 1$ and the biexciton $B$. Due to 
Coulomb interaction between electrons and holes, the energy difference 
between $B$ and $X^\pm$ is different (usually lower) to the one between 
$X^\pm$ and $G$. Excited states of the excitons as p, d, ... 
hydrogen-like states are not included.
Although in self-organized grown QD, the actual exciton eigenstates are 
usually not completely degenerated, each one being a linear combination of 
$X^+$ and $X^-$, there is always the possibility of applying an external 
magnetic field to compensate this geometric effect recovering 
the exciton basis ($X^+$, $X^-$) we are working 
with \cite{bonadeo,kulakovskii,bayer}. 
The system formed by the QD and the cavity is described by a Hamiltonian
($\hbar =1$)
\begin{eqnarray}
& & H_{S} = (\omega _C+\Delta _1 ) \left[ \mid X^+ \rangle \langle X^+ \mid 
+ \mid X^- \rangle \langle X^- \mid \right] \nonumber \\ & & + 
(2 \omega _C+\Delta _1 +\Delta _2) \mid B \rangle \langle B \mid \nonumber \\ & &
+ \sum _{J=R,L} \left[ \omega _C \left( a^\dagger _J a_J+1/2 \right) 
+ \sum _{i} q_{i,J} \left( \sigma _i a^\dagger _J + a_J \sigma 
^\dagger _i \right) \right] \nonumber \\ 
\label{hamilt} 
\end{eqnarray}
where $a_{J}^\dagger$ and $a_{J}$ are the creation and annihilation operators 
for cavity photons, of frequency $\omega _C$, with right ($J \equiv R$) 
and left ($J \equiv L$) circular polarizations. $\sigma _i$ are the set of 
four operators $\mid B \rangle \langle X^+ \mid$, $\mid X^+ \rangle \langle 
G \mid$, $\mid B \rangle \langle X^- \mid$ and $\mid X^- \rangle 
\langle G \mid$. Once again this is only an 
approximation to actual spectra in pillar or photonic crystal cavities. 
$q_{i,J}$ are the couplings between cavity modes and QD excitations. 
For simplicity, we have just written in (\ref{hamilt}) the particular case of 
degenerate excitons. The generalization to a non-degenerate case is 
obvious and we will comment below on the results in more general cases. The 
excitations energies in the QD are detuned with respect to the cavity mode
frequency: $E_{X^\pm}-E_G= \omega _C+\Delta _1$ and $E_B-E_{X^\pm}= 
\omega _C+\Delta _2$. As discussed above, the case we are interested in is 
when the detunings verify $\Delta _1 =-\Delta _2$ in order to study 
processes producing efficient pair photon emission.

The above system is not isolated from the outside world. First of all, an 
essential point, obviously necessary in any experimental situation, but not 
considered theoretically before for a four level 
system \cite{stace}, is the external excitation of the system. 
We consider an incoherent pumping (either optical 
or electrical), with rate $P$, which implies 
the lack of any upper restriction in the number of photons inside the cavity. 
Moreover, the system can emit either cavity photons with a rate $\kappa$ or 
leaky modes, with a rate $\gamma $, produced by transitions 
between the QD states. In the total Hamiltonian the pumping and emission 
processes are described by the generalization of similar terms appearing 
in the two-level case \cite{benson99,perea}. 

The magnitudes of interest for describing photon emission are obtained from 
two-times correlation functions or their Fourier transforms \cite{walls}. 
They can be calculated, by using the quantum regression theorem\cite{walls}, 
from the dynamics of the density matrix in which the degrees of freedom 
of the external reservoir have been traced out. We have computed such dynamics 
by means of a master equation within the usual rotating wave and Born-Markov 
approximations based on the fact that $\omega _C$ is much larger than any 
other energy or rate in the problem\cite{walls,cohen,scully}. Moreover, 
we consider only stationary regime with $t\to \infty$. The generalization of 
the method developed for a two-level system \cite{perea} brings to a 
master equation:
%\begin{widetext}
\begin{eqnarray}
& & \frac{d}{dt} \rho = i \left[ \rho, H_S \right] +
\sum _{J=R,L} \frac{\kappa}{2} \left( 2 a_J \rho a^{\dagger}_J 
- a^{\dagger}_J a_J \rho - \rho a^{\dagger}_J a_J  \right) 
\nonumber \\ & & +
\sum _{\begin{array} {c} i=j \\ i \oplus j \equiv 
\mid B \rangle \langle G \mid \end{array}} 
\left[ \frac{\gamma}{2}
\left( 2 \sigma_i \rho \sigma^{\dagger}_j e^{it(\omega_i-\omega_j)} 
%\right. \right. \nonumber \\ & & \left. \left. 
- \sigma^{\dagger}_i \sigma_j \rho - \rho \sigma^{\dagger}_i \sigma_j \right) 
\right. \nonumber \\ & & \left. 
+ \frac{P}{2} \left( 2 \sigma^{\dagger}_i \rho \sigma_j 
e^{it(\omega_i-\omega_j)}-\sigma_i \sigma^{\dagger}_j\rho - 
\rho \sigma_i \sigma^{\dagger}_j \right) \right] . 
\label{masterequation}
\end{eqnarray}
%\end{widetext}
Eq. (\ref{masterequation}) is represented in a basis $\mid QDS, n_R, n_L 
\rangle $ where $QDS$ is the set of the four QD states, $n_R$ and $n_L$ the 
number of right and left cavity photons respectively. This representation, 
depicted in figure \ref{fig1}, implies an infinite set of differential 
equations which we truncate at a maximum number of excitations 
(excitons plus photons) taken as large 
as necessary (typically 15). This means solving numerically, by means of 
a Runge-Kutta method, a set of a few thousands differential equations. 

The properties of the photon pairs emitted outside the cavity can be studied 
from magnitudes inside cavity\cite{walls,stace,perea}. The discussion is 
simplified by taking 
$\gamma \ll \kappa $ so that the emission of leaky modes is negligible 
compared with the rates of cavity photons escaping from the cavity. 
Then, the probabilities of detecting, outside the cavity, single and two photons
are proportional to 
\begin{eqnarray}
G_{J}^{(1)}(t,t+\tau) & =  & \langle a_{J}^\dagger
(t+\tau) a_{J}(t) \rangle , \label{G1} \\ 
G_{J,J'}^{(2)}(t,t+\tau) & = & \langle
a_{J}^\dagger(t) a_{J'}^\dagger(t+\tau) a_{J'}(t+\tau)a_{J}(t) \rangle 
\label{G2}
\end{eqnarray}
respectively (with $J \equiv R,L$). The analysis and comparison among these 
probabilities will produce the first of our main results mentioned above.

As mentioned in the introduction, the second aspect we will concentrate on 
is two-beam experiments as those performed in other systems
\cite{ou88,kwiat99,volz01,zwiller02} in which one measures linear polarizations 
which are different on each of the beams. As discussed above, in our case this 
requires a post-selection (probabilistic) process by a non-polarizing beam 
splitter separating the photon pair in two beams. After such separation, 
the experiment is similar to those \cite{ou88,kwiat99,volz01,zwiller02}
measuring coincidences related to a second order correlation 
function with no delay ($\tau=0$) 
\begin{equation}
G_{\theta }^{(2)}=\langle a_{0}^\dagger a_{\theta }^\dagger a_{\theta } 
a_{0} \rangle =\frac{G_{R,R}^{(2)}+G_{L,L}^{(2)}}{4}+G_{R,L}^{(2)} \cos ^2 \theta  
\end{equation}
where $a_{\theta }= \cos \theta (a_R +a_L)/{\sqrt 2} +i \sin \theta 
(a_R -a_L)/{\sqrt 2}$. Making $\tau =0$ in Eq. (\ref{G2}), the second 
order correlation functions take, in our basis, the form: 
\begin{eqnarray}
& G_{R,R}^{(2)}=& \sum _{QDS, n_R, n_L} n_R (n_R-1) \langle QDS, n_R, n_L \mid 
\rho \mid QDS, n_R, n_L \rangle , \label{g2rr} \\
& G_{R,L}^{(2)}=& \sum _{QDS, n_R, n_L} n_R n_L \langle QDS, n_R, n_L \mid 
\rho \mid QDS, n_R, n_L \rangle \label{g2rl} ,
\end{eqnarray}
with an expression for $G_{L,L}^{(2)}$ similar to that of $G_{R,R}^{(2)}$. 
We have taken the direction of polarization of one the beams as the 
reference because $G_{\theta}^{(2)}$ is only a function of a continuous unknown, 
the relative angle $\theta $ between the two polarizations. The visibility 
\begin{equation}
{\cal V}=\frac {\max G_{\theta }^{(2)}-\min G_{\theta }^{(2)}}
{\max G_{\theta }^{(2)}+\min G_{\theta }^{(2)}} =
\frac{2G_{R,L}^{(2)}}{2G_{R,L}^{(2)}+G_{R,R}^{(2)}+G_{L,L}^{(2)}} 
\label{visib}
\end{equation}
of the function $G_{\theta }^{(2)}$ characterizes the degree of polarization 
entanglement in the photon pair\cite{ou88,kwiat99,volz01,zwiller02}.

\section{Results}
The first case we consider is one having a set of parameters intermediate 
between the different systems where strong coupling has been 
found\cite{reithmaier,yoshie,peter}. All the couplings are taken equal 
$q_{i,J}=q=0.1 meV$, the rate of emission of leaky modes is very small 
$\gamma=0.01 meV$ and the rate of emission of cavity modes $\kappa =0.1meV$
is equal to $q$. Since in the experiments \cite{reithmaier,yoshie} 
the detunings are varied in a range larger than $q$, we take 
$\Delta_1=-\Delta_2=\Delta=0.5 meV$ as a typical value. 

Figure \ref{fig2} shows the emission spectra 
\begin{eqnarray}
S_J(\omega ) \propto 
\Re \int _0 ^\infty  d\tau e^{i\omega \tau} G^{(1)}_{J} (t,t+\tau)
\end{eqnarray}
in the stationary limit ($t\rightarrow \infty $). Results for two different 
pumping rates $P=0.05meV$ and $P=10 meV$ are given to show that a strong 
pumping introduces a strong decoherence \cite{perea} masking the features 
of the spectrum. The low pumping result shows a strong peak at the cavity 
mode frequency and a couple of satellites at the exciton and biexciton 
transition frequencies. Despite we are giving only the spectrum of cavity 
photons, satellites appear as a signature of the strong coupling regime.  

The pair emission efficiency ($\propto G^{(2)}$) compared with the single 
photon emission probability ($\propto G^{(1)}$) is usually represented by 
a parameter ($g$ in reference \cite{santori}) which in our case of 
continuous pumping becomes the second order coherence function at zero delay 
\begin{eqnarray}
g^{(2)}_{J,J'}(\tau =0)=G^{(2)}_{J,J'}(0)/G^{(1)}_{J}(0)G^{(1)}_{J'}(0).
\end{eqnarray} 
Figure \ref{fig3} shows $g^{(2)}_{R,R}$ and $g^{(2)}_{R,L}$ for values of 
$q$, $\gamma$ and $\Delta $ considered above as typical of currently 
available samples. Now $P$ and $\kappa$ are not fixed but vary along the two
axis of the figures also in experimentally accessible regimes. The results do not 
show a very rich structure and, worse than that, the best efficiency for 
emitting pairs, i.e. the highest value of $g^{(2)}_{R,L}$ is not very high. 

Since we do not get the kind of results we were looking for, we consider a 
second set of parameters not far from the previous ones: one needs either 
higher values of $q$ or lower regimes for $\kappa$, in a factor of the 
order of 2 to 5. Therefore, we force a little bit the parameters and 
consider a second case. Instead of fixing all the values in $meV$, all 
the rates and energies are given in units of $q$ having in mind that a 
possible value for this scale could be $0.2meV$ or $0.3meV$. Apart from this, 
we maintain the ratios $\Delta _1 =-\Delta _2=\Delta=5q$ and $\gamma =0.1q$,
while we must change the ratio between $\kappa$ and $q$ in order to really 
be in a different case. 
Figure \ref{fig4} shows $g^{(2)}_{R,R}$ and $g^{(2)}_{R,L}$ for this new set 
of parameters in a range of $\kappa$ and $P$ that can be reasonably expected 
to be achievable\cite{note1}. Apart from getting a structure richer than the 
one in figure \ref{fig3}, the main advantage is a significantly larger value of 
$g^{(2)}_{R,L}$. Now the emission efficiency of $RL$ pairs can be considered 
as satisfactory as compared to the emission of single photons. 

Apart from the improvement that figure \ref{fig4} represents with respect to 
figure \ref{fig3}, a general trend can be drawn for both cases. For increasing
$\kappa$ and decreasing $P$, $g^{(2)}_{R,L}$ increases monotonously while
$g^{(2)}_{R,R}$ tends to zero, a value only accessible in the quantum regime.
In the whole range of parameters, $g^{(2)}_{R,L}$ is always greater than 1. 
$R$ and $L$ photons can be distinguish from each other so that, as in 
the case of classical fields \cite{yamamoto,loudon}, 
$G^{(2)}_{R,L} \geq G^{(1)}_R G^{(1)}_L$.  
On the contrary, when the two photons have the same polarization, 
they are indistinguishable introducing the term $-1$ in the parenthesis of 
Eq. (\ref{g2rr}). This reduces $G^{(2)}_{R,R}$ and produces the quantum 
effect of having values lower than 1 for $g^{(2)}_{R,R}$, i.e. quantum 
sub-Poissonian distributions for the number of photons\cite{walls,loudon,yamamoto}. 
The continuous pumping produces an emission of pairs different to the usual 
sequential two-photon cascade in which the emission of the second photon is 
only possible after the emission of the first photon\cite{loudon}. 
  
In order to realize the significance of our results for a continuous 
pumping of a QD within a cavity, one must point out 
that available experimental results for pulsed excitation of a QD 
without any cavity and collinear polarization detection 
($\theta =0$)\cite{santori} give $g$ and $g^{(2)}$ 
in the (much smaller) range between 0.01 and 0.1. 

Our understanding of the results in figures \ref{fig3} and \ref{fig4} allows 
us to analyze the second aspect we mentioned in the introduction: the 
polarization entanglement visibility $\cal{V}$ in a experiment with different
linear polarizations at the two beams produced by a non-polarizing beam splitter. 
In order to obtain a large visibility in Eq. (\ref{visib}), one needs a small 
value of $G_{R,R}^{(2)}$. This is precisely the quantum regime 
with $g_{R,R}^{(2)}<1$ occurring for low $P$ and large $\kappa $ 
as observed in our results for $\cal{V}$ shown in figures \ref{fig5}
and \ref{fig6}. Once again, the visibility for the second set of parameters, 
shown in figure \ref{fig6}, raises up to values closer to 1 than those 
corresponding to the current samples case shown in figure \ref{fig5}.

\begin{figure}
\includegraphics [clip,height=6cm,width=8.cm]{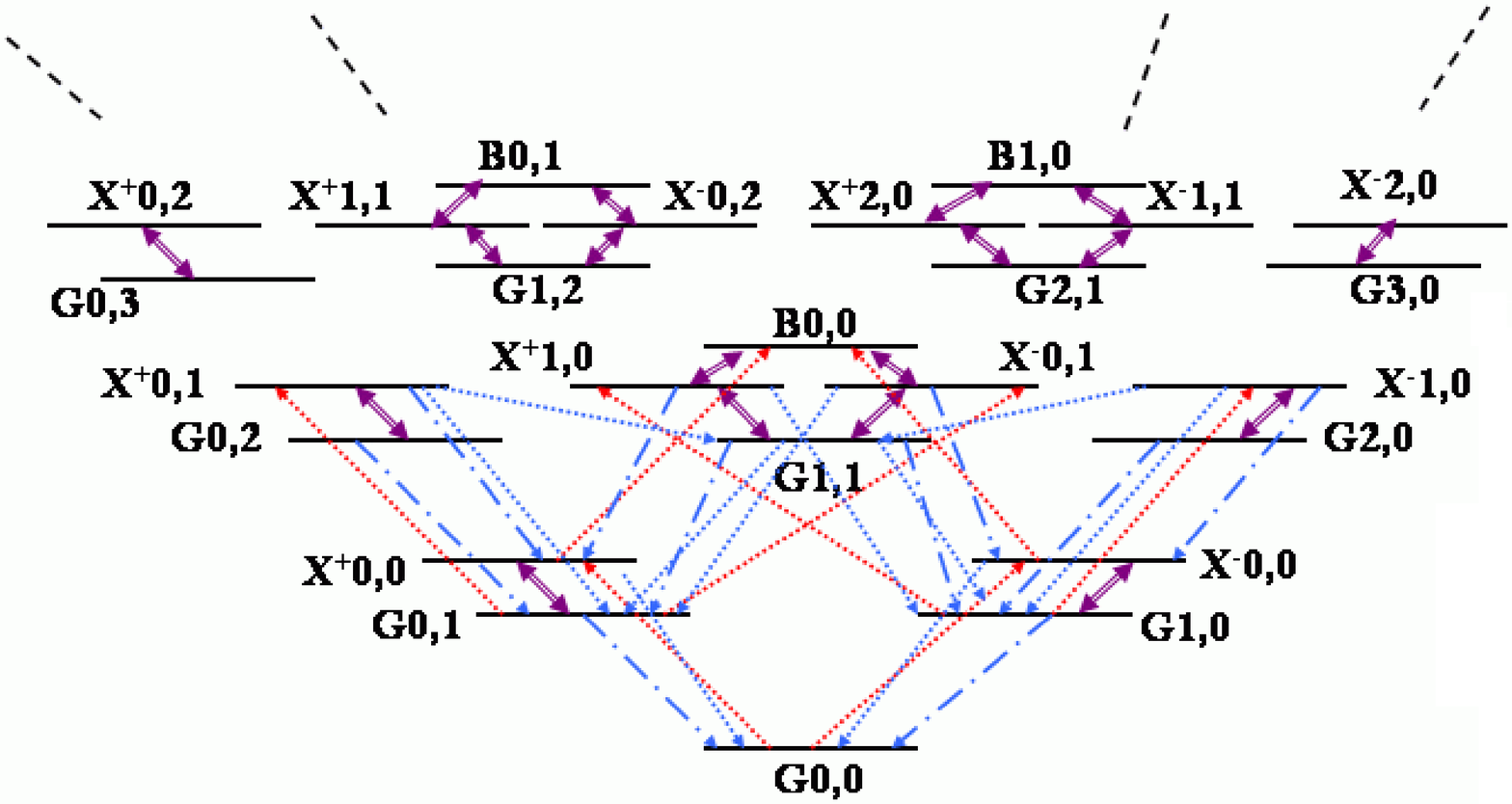}
\caption{(Color on line) Ladder of levels (black continuous lines) for a 
four-state QD coupled to a cavity mode with the two possible ($R,L$) circular 
polarizations. States are labeled as described in the text with 
$G n_R n_L $ ($n_R,n_L=0,1,2,...$). Double (purple) continuous 
lines depict the coupling $q_{i,J}$, dashed (red) lines the pumping with 
rate $P$, dotted (blue) lines the leaky modes emission with rate $\gamma $ 
and dash-dotted (blue) lines the emission of cavity modes with rate $\kappa$. 
Pumping and emission lines in the upper part of the 
diagram are not plotted in order to simplify the figure.}
\label{fig1}
\end{figure}

\begin{figure}
\includegraphics [clip,height=8cm,width=8.cm,angle=-90]{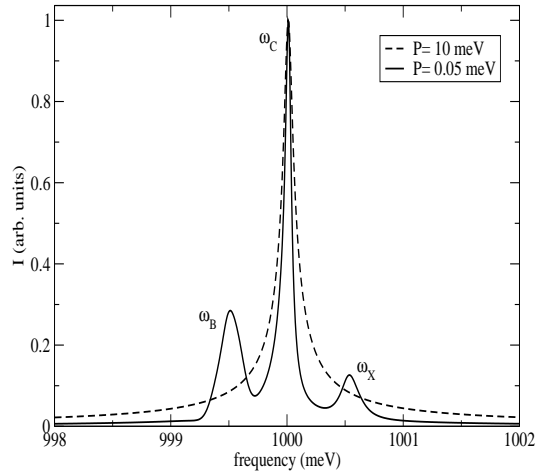}
\caption{Spectrum (in arbitrary units) of the emission of cavity photons 
for $\omega _C = 1000 $, $q=0.1$, $\Delta _1 =-\Delta _2=0.5$, $\gamma =0.01$, 
$\kappa=0.1$ and $P=0.05$ and $P=10.0$ (in dashed line) with 
all the magnitudes in meV. 
$\omega _X=\omega _C + \Delta _1$ and $\omega _B=\omega _C + \Delta _2$ 
label the frequencies for recombination of the exciton and biexciton 
respectively (see text).
}
\label{fig2}
\end{figure}

\begin{figure}
\includegraphics [clip,height=8cm,width=8.cm]{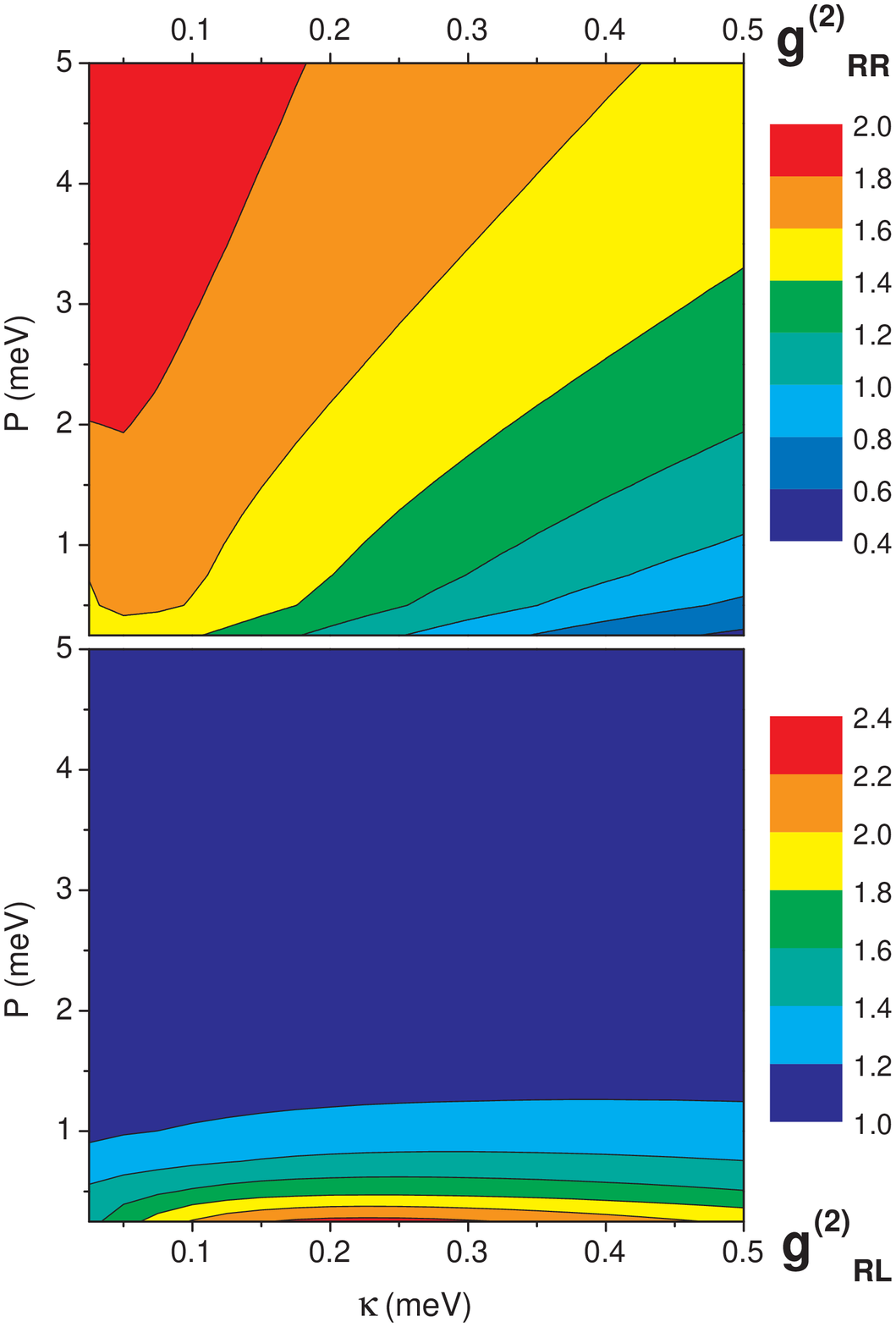}
\caption{(Color on line) Second order coherence functions $g^{(2)}_{R,R}$ 
and $g^{(2)}_{R,L}$ describing the photon pair emission efficiency ($\propto 
G^{(2)}$) compared with the single photon emission probability ($\propto 
G^{(1)}$). $q=0.1$meV, $\Delta _1=-\Delta _2=0.5$meV
and $\gamma =0.01$meV. The line where $g^{(2)}_{R,R}=1$ is labeled. 
}
\label{fig3}
\end{figure}

\begin{figure}
\includegraphics [clip,height=8cm,width=8.cm]{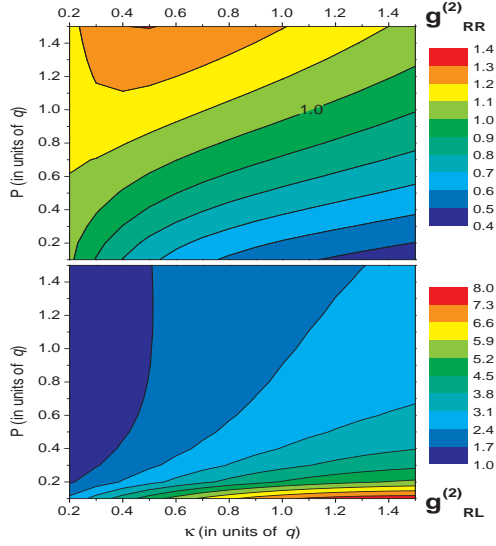}
\caption{(Color on line) Second order coherence functions $g^{(2)}_{R,R}$ 
and $g^{(2)}_{R,L}$ describing the photon pair emission efficiency ($\propto 
G^{(2)}$) compared with the single photon emission probability ($\propto 
G^{(1)}$). $\Delta _1 =-\Delta _2=5 $, 
$\gamma =0.1$ in units of $q$. The line where $g^{(2)}_{R,R}=1$ is labeled. 
}
\label{fig4}
\end{figure}

\begin{figure}
\includegraphics [clip,height=8cm,width=8.cm]{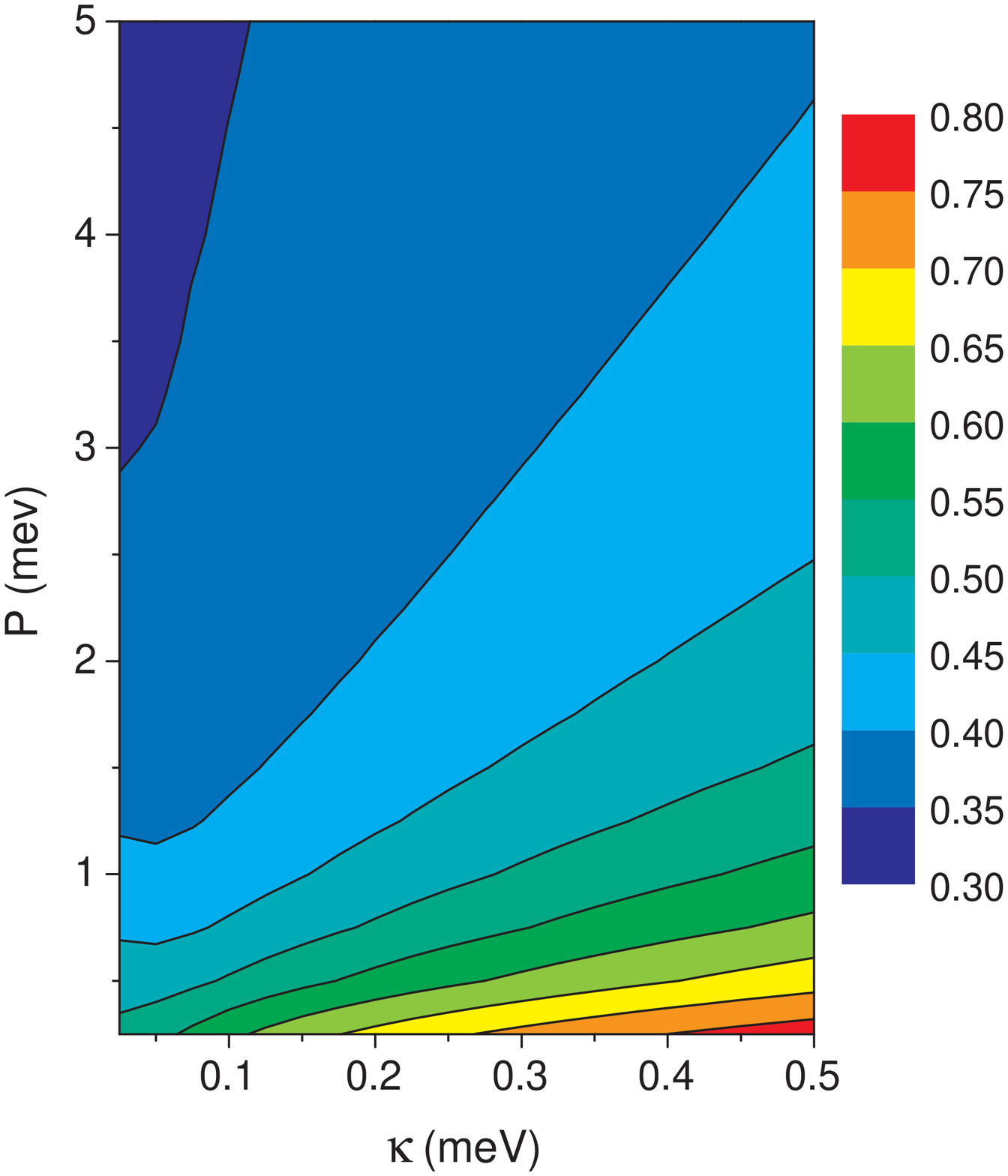}
\caption{(Color on line) Polarization entanglement visibility $\cal{V}$ for a 
pair of photons with two different linear polarizations. $q=0.1$meV, $\Delta _1 =
-\Delta _2=0.5 $meV, $\gamma =0.01$meV.
}
\label{fig5}
\end{figure}

\begin{figure}
\includegraphics [clip,height=8cm,width=8.cm]{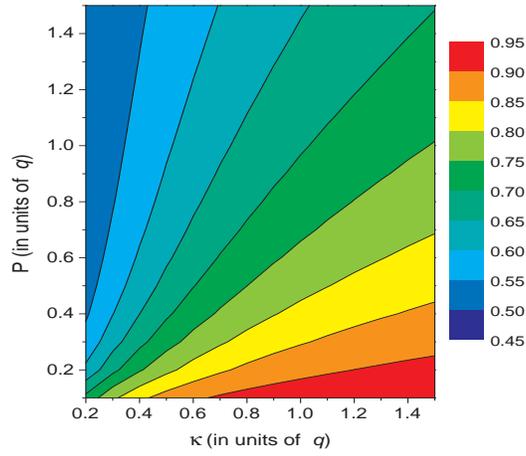}
\caption{(Color on line) Polarization entanglement visibility $\cal{V}$ for a 
pair of photons with two different linear polarizations. $\Delta _1 =
-\Delta _2=5 $, $\gamma =0.1$ in units of $q$.
}
\label{fig6}
\end{figure}

Let us try to understand the reason why our system seems to be less efficient 
than expected concerning to the emission of a photon $RL$ pair (except for  
the regime with low $P$ and large $\kappa $ in figure \ref{fig6}). The actual 
situation is not well described by a simple picture considering that 
emission of one of these $RL$ pairs is significantly improved by the double 
resonance condition $\Delta _1 =-\Delta _2$. Figure \ref{fig1} shows that, 
in our basis, apart from the ground state there are two types of subsets 
with either two or four states, for instance 
the lowest subset with four states contains $\mid B00 \rangle $, 
$\mid X^+ 01 \rangle $, $\mid X^- 10 \rangle $ and $\mid G 11 \rangle $. 
Inside one subset, there are not transitions contributing to the correlation 
functions $G^{(1)}$ and $G^{(2)}$. Therefore the processes occurring inside 
each subset are not as important as the simple picture would imply. On the 
contrary, the main physics behind photon emission is coming from transitions 
among different subsets as depicted in figure \ref{fig1}.

Finally, we must stress that we have repeated our analysis for 
QD's with different characteristics, in particular for two cases:
{\it i)} when the two excitons $X^+$ and $X^-$ are not degenerate 
and {\it ii)} when there is not perfect (but close to) double resonance 
$\Delta _1 \neq -\Delta _2$. We have found quantitative changes, but in all 
the cases, the results shown here remain qualitatively valid. 

One can conclude that, in spite of not being as efficient as expected for 
emitting $RL$ pairs of photons, the system shows a rather rich and complex 
behavior including both quantum and classical regimes. Our results suggest 
that, even though currently available samples are not in the best regime, 
they are close enough to expect that one can achieve a regime of efficient 
emission of photon pairs. In the quantum regime, a post-selection procedure 
would allow to perform a two beams experiment with different linear 
polarizations in which a large polarization entanglement visibility could 
be achieved.

\section{Acknowledgments}
We are indebted to Filippo Troiani for very helpful discussions.
This work was supported in part by MCYT of Spain under contract No.
MAT2002-00139, CAM under Contract No. GR/MAT/0099/2004 and European
Community within the RTN COLLECT.

%%%%%%%%%%%%%%%%%%%%%%%%%%%%%%%%%%%%%%%%%%%%%%%

\bibliography{pevv2}

\begin{thebibliography}{28}
\bibitem[{not()}]{note1}
\bibinfo{note}{If one takes $q=0.1$meV as in figure 3, the range of parameters
  of figure 4 is below and to the left of the range of figure 3}.
\end{thebibliography}

%%%%%%%%%%%%%%%%%%%%%%%%%%%%%%%%%%%%%%%%%%%%%%

\end{document}